# Physics instructors' knowledge and use of active learning has increased over the last decade but most still lecture too much.

Melissa Dancy, The Evaluation Center, Western Michigan University
Charles Henderson, Dept. of Physics and Mallinson Institute for Science Ed., Western Michigan University
Nanah Apkarian, School of Mathematical and Statistical Sciences, Arizona State University
Estrella Johnson, Department of Mathematics, Virginia Tech University
Marilyne Stains, Department of Chemistry, University of Virginia
Jeffrey R. Raker, Department of Chemistry, University of South Florida
Alexandra Lau, American Physics Society

## Abstract

A survey of 722 physics faculty conducted in 2008 found that many physics instructors had knowledge of research-based instructional strategies (RBISs), were interested in using more, but often discontinued use after trying. Considerable effort has been made during the decade following 2008 to develop and disseminate RBISs in physics as well as change the culture within the physics community to value RBISs use and other forms of student-centered instruction. This paper uses data from a 2019 survey of 1176 physics instructors to understand the current state of RBIS use in college-level introductory physics, thus allowing us to better understand some of the impacts of these efforts on physics instruction. Results show that self-reported knowledge and use of RBISs has increased considerably and discontinuation is now relatively low. However, although the percentage of time lecturing is less than 10 years ago, many instructors still engage in substantial lecturing. Relatedly, we find that the majority of RBIS use centers on pedagogies designed to supplement traditional lecture rather than pedagogies designed to supplement an active learning classroom. This suggests that the PER community and beyond has done well promoting knowledge about RBISs and inspiring instructors to try RBISs in their courses. But, there is still room to improve. We recommend that change agents focus on supporting instructors to increase the percent  of class time in active learning and to implement higher impact strategies

## I - Introduction

A survey of physics faculty conducted in 2008 found that many physics instructors had knowledge of research-based instructional strategies (RBISs) [1], were interested in using more RBISs in their courses [2], but often discontinued use after trying one or more RBISs [3].  Considerable effort has been made during the decade following 2008 to develop and disseminate RBISs in physics as well as to change the culture within the physics community to value RBIS use and other forms of student-centered instruction [4]. This paper uses data from a 2019 survey to understand some of the impacts of these efforts on physics instruction.

# Efforts to increase the adoption and sustained use of research based instructional strategies in physics

Rogers [5] proposes that adoption of a new practice occurs over time in a series of five decision-making stages: Knowledge about the innovation, Persuasion about the benefits of the innovation, Decision to use the innovation, Implementation of the innovation, and Confirmation of continued implementation of the innovation. In our 2008 study of physics instructors in the US, we found that knowledge about RBISs and persuasion about the benefits of RBISs were high [1,2]. Many instructors also had made the decision to use a RBIS and tried implementing it [3]. However, we also found that discontinuation was high, with about ⅓ of the instructors who tried to use a RBIS discontinuing the use of that RBIS and all other RBISs that we asked about [3]. Thus, we recommended that more attention be paid to supporting instructors during implementation so that they could implement successfully and not discontinue [3].

Whether due to these recommendations or for other reasons, in the decade after 2008, change agents have placed additional emphasis on providing support during implementation along with development and dissemination activities. Perhaps the most far-reaching of these efforts is the Physics and Astronomy New Faculty Workshop (NFW). The NFW is important because approximately 40% of all new physics faculty in the US attend the workshop each year [6]. In addition to the in-person workshop, beginning in 2015, NFW attendees were given the opportunity to participate in a Faculty Online Learning Community (FOLC) [7-10]. The FOLC experience included virtual meetings with a small group of NFW attendees and a facilitator approximately every other week for approximately one year after the NFW. The goal of the FOLC was to provide support for faculty as they attempt to implement RBISs. Discussions are focused on implementation difficulties and successes. Much of the support is provided by the other FOLC participants, but outside experts are also invited if needed. In addition to FOLC use as part of the NFW, FOLCs have been used in other settings to implement RBISs in physics [11,12].

Another example of efforts to provide more support for instructors implementing RBISs is the Carl Wieman Science Education Initiative (CWSEI) at the University of British Columbia [13]. In the CWSEI model, science education specialists are hired to work with faculty to transform courses taught by those faculty [14]. These specialists are typically PhDs with special training in RBISs and discipline-based education research. Wieman et al. [13] found that, of the 70 faculty who implemented RBISs as part of the CWSEI, only one discontinued use. They attribute this high continuation to the support offered by the science education specialists. The specialists help faculty customize the RBIS and are also available to help troubleshoot implementation difficulties. As the authors note "Having a knowledgeable person who can minimize the initial challenges of implementation and ensure that RBISs are successful and well received by students when first implemented is an enormous step towards encouraging faculty to embrace the use of RBISs."[13, pg. 3]

## Cultural Changes - RBIS use becoming more acceptable

In addition to more focus on providing support to instructors during initial use of RBISs, there is also evidence that the expectations for physics teaching in the US have been changing. That is, many higher education institutions and physics departments within them, now value the use of RBIS and encourage their faculty to adopt these strategies. For example, the recently-published Effective Practices for Physics Departments guide was developed by the American Physical Society and advocates for the use of research-based instructional practices and inclusive pedagogy in physics courses [15].

Given these changes to the way that advocates of RBIS use in physics have focused more on supporting users, as well as the cultural changes in the expectations for physics teaching it is important to understand whether the biggest problem in the improvement of physics teaching in higher education is still the lack of support. And, if not, what new barriers exist that will be important to address.

# II - Data Collection

The goal of this paper is to compare results of the 2008 survey of physics faculty [1-3] with a more recent 2019 survey of physics faculty. Both surveys were focused on instructors teaching introductory-level physics in the US. We describe each survey in the following sections.

## 2008 Survey (n=722)

The 2008 survey was developed by two of the authors (Dancy and Henderson) in consultation with researchers at the American Institute of Physics Statistical Research Center. Questions focused on their teaching situation, experience and attitudes toward teaching innovations, their instructional goals and practices and demographic information. Faculty were eligible for the survey if they had taught an introductory quantitative course in the last two years and were full-time or permanent employees (i.e., faculty who were part-time, temporary employees were not eligible for the survey).

The survey was administered in Fall 2008 by the American Institute of Physics Statistical Research Center (SRC). Sampling was done at three types of institutions: (1) two-year colleges (TYC), (2) four-year colleges that offer a physics bachelor's degree as the highest physics degree (BA), and (3) four-year colleges that offer a graduate degree in physics (GRAD). 722 usable responses were collected from instructors at 345 different institutions. The overall response rate was 50.3%

Further details on the 2008 survey as well as results of analysis can be found elsewhere [1-3].

## 2019 Survey (n=1176)

The 2019 survey was designed, in part, as a follow-up to the 2008 survey. Similar to the 2008 survey, the sample included postsecondary instructors who had taught introductory physics courses, not entirely online, in the previous two years, at two-year colleges (TYC), four-year colleges without graduate degrees in physics (BA), and four-year colleges that offer a graduate degree in physics (GRAD). The 2019 survey also sampled instructors in chemistry and mathematics in addition to physics. Only the physics data is presented here; other findings and more details on survey distribution are discussed elsewhere [16, 17].

The 2019 survey was developed by six of the authors (MD, CH, EJ, NA, MS, JR) for this project. The full survey covered five main topics: (1) course context and details; (2) instructional practice; (3) awareness and usage of active learning instruction; (4) perceptions, beliefs, and attitudes related to students, learning, and departmental context; (5) personal demographics and experience. A web-based version of the instrument was built and distributed in partnership with the American Institute of Physics Statistical Research Center. Stratified random sampling was done by institution based on institution type, with the goal of developing a representative sample of institution types. Invitations were sent to over 18,000 individuals identified from publicly available information (e.g., institution website) and communication with department chairs by members of the American Institute of Physics Statistical Research Center. The full survey was answered by 3,769 instructors of which 1,176 were collected from physics instructors at 565 different institutions.

# III - Respondent Demographics

Demographics of respondents from the two surveys are shown below in Table 1. All demographic identities are self reported except for institution type. Demographics of respondents were similar across both surveys. Therefore any differences in results are most likely the result of time and not of a different population sampled.

|  | 2008 | 2019 |
|---|---|---|
| **Type of Institution** | | |
| Two Year College | 25.8% | 22.2% |
| Undergraduate Program | 35.3% | 31.7% |
| Graduate Program in Physics | 38.9% | 46.1% |
| **Academic Rank** | | |
| Lecturer/Instructor/Adjunct | 14.3% | 21.5% |
| Assistant Professor | 20.8% | 16.6% |
| Associate Professor | 24.2% | 22.9% |
| Full Professor | 35.6% | 36.7% |
| Other Rank | 5% | 2.2% |
| **Gender** | | |
| Female | 17% | 21.3% |
| Male | 83% | 78.6% |

| Other Gender | n/a | <1% |
|---|---|---|
| **Semesters Taught** | | |
| 1-4 Semesters | 15% | 5.1% |
| 5-10 Semesters | 20% | 13.3% |
| >10 Semesters | 65% | 81.6% |

Table 1 - Demographics of survey respondents in 2008 and 2019.

# IV - Results

In this paper we present three comparisons between the 2008 and 2019 surveys. First, we look at how self-reported knowledge and use of specific Research-Based Instructional Strategies (RBISs) has changed over time. We then compare how overall reports of active learning have changed. Finally, we compare where respondents are in Rogers' innovation decision process in the two surveys.

## Knowledge and use of specific RBIS have increased significantly between 2008 and 2019

In both the 2008 survey and the 2019 survey, we presented respondents with a list of specific and common RBISs. The 2008 survey asked about 24 RBISs and the 2019 survey asked about 14 with 8 RBISs overlapping both surveys (see Figure 1 for the 8 overlapping RBISs) . In both surveys respondents were presented with an RBIS along with a description of the RBIS and then asked to describe their familiarity and use of each RBIS by selecting from 5 options.  These options were slightly different in the two surveys. Table 2 provides details on the answer choices and how they were organized into knowledge and use categories.

| Categorization in analysis | Roger's (1995) stages | 2008 Answer Choice | 2019 Answer Choice |
|---|---|---|---|
| No Knowledge | Knowledge | "I have never heard of it." | "I have never heard of this." |
| Knowledgeable Non-User | Persuasion Decision | "I've heard the name, but do not know much else about it." or "I am familiar with it, but have never used it" | "I know the name, but not much more" or "I know about this, but have never used it in this course." |
| Former User | Implementation | "I have used all or part of it in the past." | "I have tried it in this course, but no longer use it." |

| Current User | Confirmation | "I currently use all or part of it." | "I currently use it in this course to some extent." |

*Table 2 - Categorization scheme for specific RBIS knowledge and use. Survey respondents were asked to think about a specific introductory course they taught as the primary instructor in order to answer the question.*

Of these 14 strategies 8 were the same as strategies asked about in the 2008 survey. Figure 1 compares levels of knowledge and use reported in 2008 and 2019 for the 8 strategies that were on both surveys.

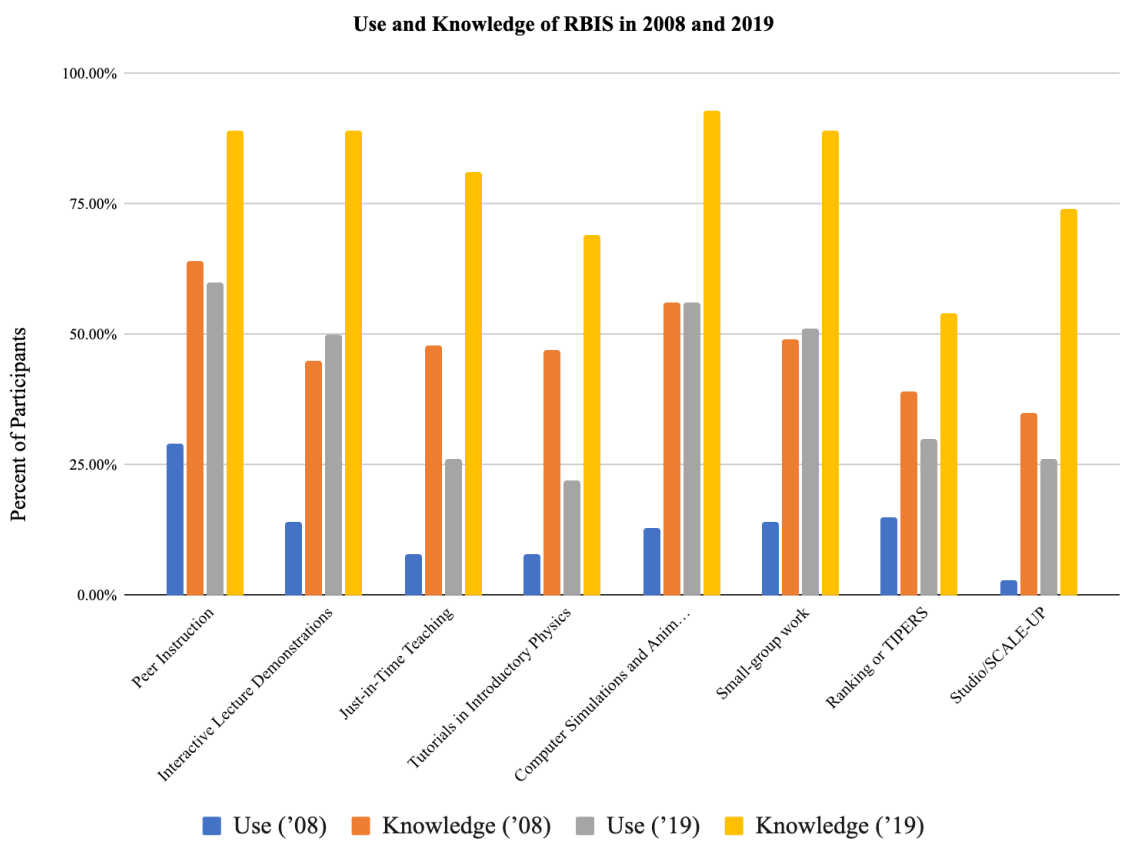

*Figure 1 - Comparison of reported knowledge and use of specific RBISs in 2008 and 2019. Respondents are considered to have knowledge if they were classified as either a knowledgeable non-user or a former user. If a respondent indicated they were a user, they were also counted as being knowledgeable about the RBIS.*

For every strategy, both knowledge and use levels have greatly increased over the 10 years between surveys. Notably, knowledge levels are reaching saturation (i.e nearly all instructors are aware of RBIS's) and use levels are reported to be high (i.e. a majority of instructors report use).

# The amount of time instructors report lecturing decreased significantly between 2008 and 2019 but is still high.

In addition to asking respondents about their knowledge and use of specific RBIS we also asked about the time they utilized particular types of general classroom activities in their teaching. Here we compare the relative time instructors spend lecturing vs. utilizing active learning techniques. The questions were asked differently on the two surveys as described below.

2008 Survey: Respondents were asked to respond to the question "In the 'lecture portion' of your introductory course, please estimate the percentage of class time spent on student activities, questions, and discussion." They could enter any number into a textbox.

2019 Survey: Respondents were asked, "What proportion of time during regular class meetings do students spend listening to the instructor lecture or solve problems." Answer choices were provided in increments of 5%. For the purposes of comparison, we take the percent of time spent on student activities to be the time not reported to be spent listening to the instructor lecture or solve problems.

Figure 2 shows a comparison of responses to these two questions on both surveys. Consistent with reports of increasing use of RBISs, we find that reports of general active learning use have significantly increased. Of note is that the number of instructors reporting using lecture almost exclusively has fallen by about half over the years. This finding is very encouraging. However, more work is needed to support instructors in decreasing the time they spend in lecture.

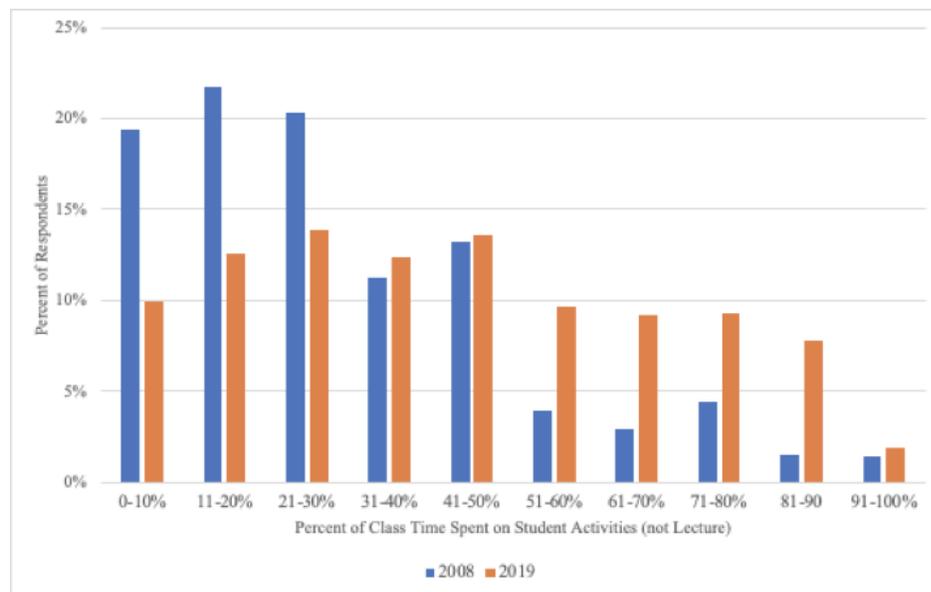

Figure 2 - Comparison of reported time spent on student activities.

Theobald et. al. [18] report on a meta-analysis of studies of undergraduate STEM classrooms comparing exam scores and passing rates of racially under and over represented students experiencing an active learning or traditional lecturing class format. They found that high use of active learning significantly decreased achievement gaps, where high use of active learning was defined as using active learning for more than ⅔ (63%) of class time. We find that for physics instructors today, less than 30% report meeting this threshold. While the large increases in reported time in active learning are encouraging, and almost all faculty report knowing about and using at least some active learning techniques, the majority are still not using active learning at sufficiently high levels to reach the best results.

## Instructors still utilize mostly RBISs that supplement lectures rather than strategies that focus the course on active learning.

While the findings reported above show that significant progress has been made in the quality of introductory physics instruction in US higher education, there is still substantial room for improvement. One of the things that we know from the literature is that, while use of any RBIS is better than a traditional lecture class [19,20], the more that a RBIS incorporates student-centered activities over instructor-activities, the higher the levels of student learning tend to be [18, 21-25]. Thus, while one of the strengths of Peer Instruction is that it is easily incorporated into a traditional lecture-based physics course, one of the weaknesses is that it is not as large a departure from a traditional lecture-based course as RBISs such as studio-style instruction that require fundamental changes to the course structure.

Data from the 2019 survey for all RBISs asked about is presented in Tables 3 & 4. In Table 3 we rank the 14 RBISs queried based on the percent of respondents who indicated knowledge of the RBIS (i.e. they indicated knowledge regardless of reported use). In Table 4 we rank the RBISs based on the percent of respondents indicating they are current users of the RBIS. Table 4 shows that there are four RBISs used by more than 50% of faculty (Peer Instruction, Computer Simulations and Animations, Concept Inventories, and Small-Group Work). All of these are relatively easy to incorporate into a traditional lecture setting. All of the RBISs that require significant departure from a traditional lecture class and, thus are likely to result in stronger student learning, are used by less than ⅓ of faculty. These include Flipped Classroom, Studio/SCALE-UP, and Just-in-Time Teaching.

Table 3: Percent of respondents reporting knowledge of the RBIS in 2019, broken down by institution type. The RBISs are ordered from the strategy with highest level of knowledge to least. Knowledge is defined as respondent reporting being familiar with the strategy, having used it in the past, or are currently using it.

| **Reported Knowledge of Common RBISs** | | | | |
|---|---|---|---|---|
| RBIS | All Institutions | TYC | PUI | UNI |
| Comp Sim, Anim | 93% | 94.5% | 94.9% | 90.3% |
| Flipped Classroom | 91.9 | 88.2 | 94.7 | 91.3 |
| Peer Instruction | 89.2 | 81.1 | 93 | 89.9 |
| Small-Group Work | 88.9 | 84.8 | 92 | 88.2 |
| Interactive Lecture Demos | 88.5 | 85.3 | 93 | 86.1 |
| Concept Inventories | 87.3 | 76.6 | 94.2 | 86.6 |
| Just-in-Time Teaching | 81.4 | 72.3 | 86.3 | 81.7 |
| Think-Pair-Share | 80.9 | 76.2 | 86.3 | 78.3 |
| Studio/SCALE-UP | 73.6 | 66 | 76.9 | 74.5 |
| Peer-Led Team Learning | 69.9 | 60.6 | 73.1 | 72 |
| Tutorials Intro Physics | 68.8 | 63.8 | 75.5 | 65.1 |
| Concept Maps | 65.3 | 67.6 | 70 | 59.5 |
| Peer-Rev Sci Writing | 57.2 | 55 | 62.6 | 53.2 |
| Ranking or TIPERs | 54.3 | 61.3 | 57.8 | 47.1 |

Table 4: Percent of respondents reporting current use of the RBIS in 2019, broken down by institution type. The RBISs are ordered from the strategy with highest level of use across all institution types to least.

**Reported Use of Common RBISs**

| RBIS | All Institutions | TYC | PUI | UNI |
|---|---|---|---|---|
| Peer Instruction | 60.2 | 50.4 | 67.2 | 58.9 |
| Comp Sim, Anim | 56.3 | 61 | 59.8 | 50.5 |
| Concept Inventories | 52.8 | 45.2 | 60 | 50.2 |
| Small-Group Work | 51.4 | 54.9 | 53.7 | 47.4 |
| Interactive Lecture Demos | 49.5 | 50.4 | 53.3 | 45.4 |
| Think-Pair-Share | 46.7 | 43 | 55.5 | 40.4 |
| Flipped Classroom | 31.7 | 28.3 | 31.6 | 33.6 |
| Ranking or TIPERs | 29.5 | 37 | 28.9 | 26.1 |
| Studio/SCALE-UP | 25.7 | 31.5 | 26.5 | 21.8 |
| Just-in-Time Teaching | 25.5 | 17.6 | 28.7 | 26.8 |
| PLTL | 25.3 | 16.1 | 25.7 | 29.8 |
| Tutorials Intro Physics | | 21.8 | 22.6 | 23.3 | 20.1 |
| Concept Maps | | 15.8 | 21.8 | 14.5 | 13.9 |
| Peer-Rev Sci Writing | | 6.1 | 9.7 | 6.1 | 4.1 |

As discussed previously, while most instructors are using some active learning, they are not using it at sufficiently high levels to achieve important outcomes. The reliance on techniques designed to supplement traditional lecture is likely a limiting factor in increasing overall use of active learning. The amount of class time spent in active learning can likely be increased by promoting the use of strategies designed to support a primarily active learning classroom rather than strategies designed to add active learning to a traditional lecture based course.

## Discontinuation is no longer a big problem

As noted earlier, one of the major findings from the 2008 survey is that a large percentage of instructors who start using a RBIS end up discontinuing use. Note that for the purpose of our analyses, when we say that a respondent discontinued use of RBISs, we mean that they reported trying one or more RBIS, and now report not using any RBISs. Figures 3 and 4 show the level of knowledge and use for the two surveys. As shown in the figures, the percentage of instructors with no knowledge of any RBIS fell from 12% in 2008 to 2% in 2019. The percentage of instructors who have not tried a RBIS fell from 16% in 2008 to 8% in 2019. The percentage of instructors who had tried 1 or more RBISs increased from 72% in 2008 to 90% in 2019. In 2008, about ⅓ (23%/72% = 32%) of those who tried ended up discontinuing use of all RBISs. In 2019, discontinuation is a very low 5% of those who tried a RBIS.

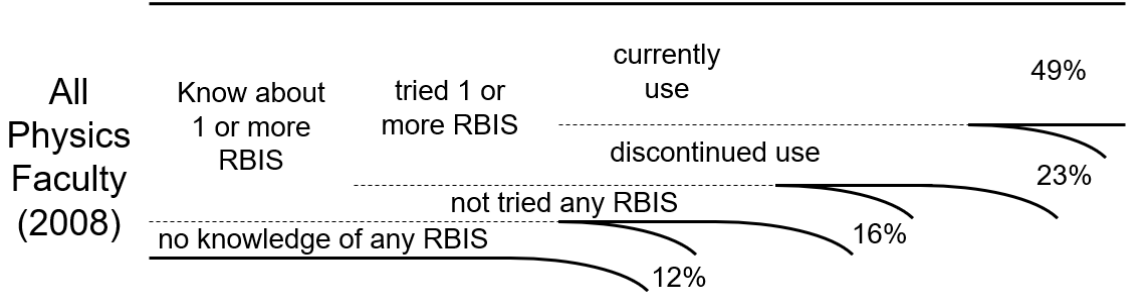

Figure 3: Classification of respondents by level of knowledge and use in 2008.

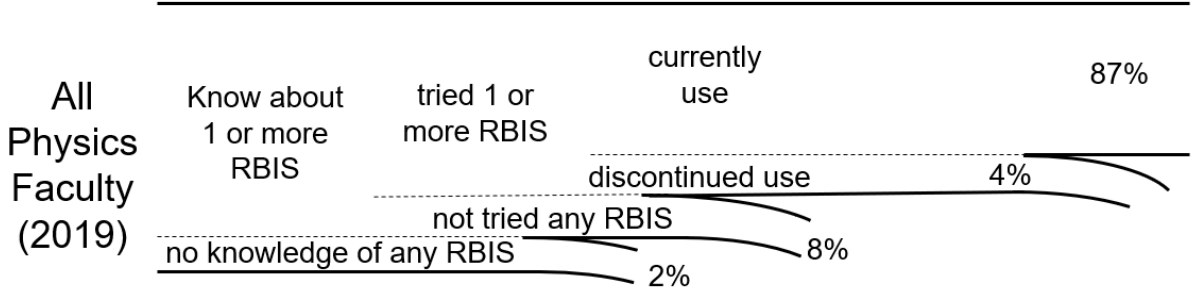

Figure 4: Classification of respondents by level of knowledge and use in 2019.

Survey results from 2008 indicated that the biggest loss point was in instructors who had tried RBISs but discontinued use. Encouragingly today discontinuing use is no longer a significant problem. Data now indicates that the main focus for change agents should be to support the 87% of instructors who are using RBISs to implement robustly and with positive impact.

## Summary/Discussion

In this paper we have examined results of a national survey of physics instructors to identify their level of knowledge and use of Research Based Instructional Strategies (RBISs). Compared to a similar survey conducted in 2008, we found that knowledge and use are now much higher. Knowledge is currently close to 100% for many RBISs, and 87% of instructors (compared to 49% in 2008) now report using one or more RBIS. Perhaps more importantly, we also found that discontinuation of a RBIS after trying fell from 32% in 2008 to 5% in 2019. These are all very promising results and suggest that work by members of the physics community to promote high quality undergraduate instruction are having a positive impact.

This study also identified that additional work is still needed. The most commonly used RBISs are the ones that are the smallest departure from lecture-based courses. While this makes them easy to implement in many situations, RBISs that are more of a departure from a traditional lecture course (such as SCALE-UP or flipped classrooms) are capable of producing stronger student outcomes. While use of these more robust active learning techniques has increased over the last decade along with lecture-based ones, their use is still not the norm. Additionally, we find that while most faculty report making use of active learning, the overall percent of class time devoted to lecture is still high resulting in limited impact of the active learning used.

Our findings suggest that the physics education community needs to shift its energy from showing that RBISs are better than traditional lecture-based courses and encouraging faculty to dip their toes into active learning, to documenting the advantages of some RBISs over others. There also continues to be a need to support instructors in shifting more of class time to active learning. **In short, change agents should focus efforts on motivating and supporting instructors to use more RBISs that are not primarily supplements to lecture (i.e. studio or SCALE-UP) and increase the time spent in active learning to be greater than ⅔ of class time.**

## Acknowledgements

This material is based upon work supported by the National Science Foundation under Grant No. (Grants 1726042, 1726281, 1726126, 1726328, & 1726379). Any opinions, findings, and conclusions or recommendations expressed in this material are those of the authors and do not necessarily reflect the views of the National Science Foundation. We would like to acknowledge and thank the participants in our study for their time and effort telling us about their instructional practices and context. Similarly, the members of the American Institute of Physics Statistical Research Center for their help gathering the survey data for this study.